\documentclass[journal,10pt,english,a4paper]{IEEEtran}

\newcommand{\figref}[1]{figure~\ref{#1}}
\newcommand{\abs}[1]{\left\lvert#1\right\rvert}

\newcommand{\iid}{i.i.d.~}

\newcommand{\mui}[1]{I\left(#1\right)}
\newcommand{\den}[1]{h\left(#1\right)}

\newcommand{\rgauss}[2]{\mathcal{N}\left( #1, #2 \right)}

\newcommand{\snr}{{\text{SNR}}}
\newcommand{\snre}{{\text{SNR}_{\text{Eve}}}}
\newcommand{\snra}{{\text{SNR}_{\text{Alice}}}}
\newcommand{\snrb}{{\text{SNR}_{\text{Bob}}}}

\renewcommand{\Re}{\text{Re}}

\usepackage{graphicx}
\usepackage{lmodern} 
\usepackage{amsmath}
\usepackage[normalem]{ulem} 
\usepackage{subcaption}
\usepackage{cite} 

\usepackage{hyperref}
\usepackage{cleveref} 

\renewcommand{\j}{\mathrm{j}}

\graphicspath{{./figures/}}
\DeclareGraphicsExtensions{.pdf} 

\begin{document}
\title{Secret-key generation from wireless channels: Mind the reflections
\thanks{This work was funded by the Federal Ministry of Education and Research (BMBF) of the Federal Republic of Germany (F\"orderkennzeichen 16 KIS 0030, Prophylaxe). The authors alone are responsible for the content of the paper. }}

\author{Hendrik Vogt, Aydin Sezgin\\Ruhr-Universit\"at Bochum, Germany\\
hendrik.vogt@rub.de, aydin.sezgin@rub.de}
\maketitle

\begin{abstract}
Secret-key generation in a wireless environment exploiting the randomness and reciprocity of the channel gains is considered.
A new channel model is proposed which takes into account the effect of reflections (or re-radiations) from receive antenna elements, thus capturing an physical property of practical antennas.
It turns out that the reflections have a deteriorating effect on the achievable secret-key rate between the legitimate nodes at high signal-to-noise-power-ratio ($\mathsf{SNR}$).
 The insights provide guidelines in the design and operation of communication systems using the properties of the wireless channel to prevent eavesdropping.
\end{abstract}
\section{Introduction}\label{introduction}
Secret-key generation based on the randomness of the wireless channel has gained much attention recently, since it provides a low-cost and yet effective alternative to existing higher layer approaches. The basic concept relies on certain properties of channel statistics, especially on the fact that spatial correlation of the channel decreases rapidly with distance \cite{src:dent1993}.
First results on exploiting dependent (or common) source of randomness to generate a secret key have been provided by Maurer\cite{src:maurer1993} and Ahlswede and Csiszar\cite{src:ahlswede1993}. The largest achievable key rate (in bits per source symbol) without leaking any information to a possible eavesdropper is termed the secret-key capacity.  In wireless communication, a possible source for generating secret keys is the wireless channel itself. Thus, different approaches have been proposed utilizing the characteristics of the channel to generate secret keys. There have been many publications regarding the theoretical and practical aspects of secret-key generation such as \cite{src:prabhakaran2012,src:mathur2008,src:bloch2008,src:wallace2009k,src:premnath2013,src:ye2010}. Recently, physical experiments have been conducted providing an estimate of the achievable secret key rate in indoor environments\cite{src:pierrot2013}. It was shown that the characteristics of the channel essential for generating secure keys, such as reciprocity between legitimate nodes and the spatial decorrelation, are not always met in practice. It was also shown recently that the coherence time of the channel plays a significant role in obtaining secret keys\cite{src:khisti2012}, and that additional measures have to be taken into account to ensure secure communication. 
Based on those results, one might argue that channel models traditionally used for communications without secrecy constraints might result in too optimistic outcomes when used for secure communication analysis. If too simplistic models are utilized to estimate the achievable secret-key rates, they do not fully consider the underlying physics relevant for secret key generation, which is the focus of this work.

From prior work, it is clear that several pitfalls occur in practical realizations. 
The authors of \cite{src:chou2010} investigate the case when the wireless channel exhibits a sparsity pattern that increases the eavesdropper's ability to observe the main channel. In another work \cite{src:dottling2011} it is shown that antennas scatter a significant amount of power. This power might be intercepted by an eavesdropper, who can use the reflected signal to draw conclusions about the current channel state between the legitimate users. Furthermore, the authors present a method how the eavesdropper can reconstruct the impact of simple environments on the channel state.

Thus, a new channel model is proposed which captures the phenomenon of reflection occurring at any real world antenna. It turns out that, by taking the reflection into account, the achievable secret-key rate is decreasing as the signal-to-noise ratio at the eavesdropper is at moderate to high levels.  Therefore there is a optimal signal-to-noise ratio to maximize the achievable secret-key rate, which has to be taken into account when designing systems.

The paper is organized as follows: In section \ref{sec:system_model}, the new system model is introduced. Section \ref{sec:sources_reflection} outlines the physical effects from antenna theory that establish the power reflection. In section \ref{sec:secret_key_rates}, the general bounds on secret-key rates are formulated, while in sections \ref{sec:rates_without_csie} and \ref{sec:rates_with_csie} the bounds are computed and evaluated for special cases. Section \ref{sec:conclusion} concludes the paper.
\section{System model} 
\label{sec:system_model}
\begin{figure}
\centering
\begin{subfigure}[htb]{0.45\textwidth}
	\centering 
	\includegraphics{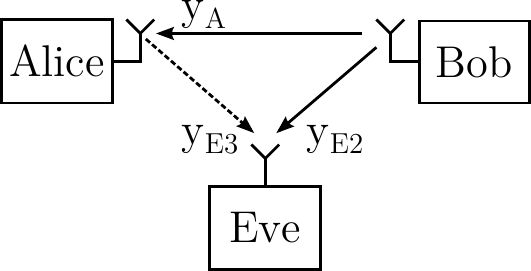}
	\caption{Training signal sent by Bob.}
	\label{fig:model_reflect1}
\end{subfigure}
~
\begin{subfigure}[htb]{0.45\textwidth}
	\centering 
	\includegraphics{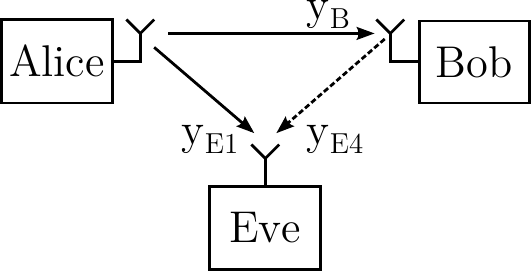}
	\caption{Training signal sent by Alice.}
	\label{fig:model_reflect2}
\end{subfigure}
\caption{Illustration of exchanged signals at the users.}
\label{fig:model_reflect}
\end{figure}

When the training signals are exchanged at the legitimate users, four signals can be defined as follows:
\begin{align}
\label{eq:observations_wo_refla}
y_A & = h_{BA}x+z_A \\
\label{eq:observations_wo_reflb}
y_B & = h_{AB}x+z_B \\
\label{eq:observations_wo_refl1}
y_{E1} & = h_{AE}x+z_{E1} \\
\label{eq:observations_wo_refl2}
y_{E2} & = h_{BE}x+z_{E2}.
\end{align}
The overall communication setup is shown in \figref{fig:model_reflect}. The legitimate users Alice and Bob exchange training signals in order to estimate correlated channel coefficients. The deterministic training symbol $x$ is set to $x=1$ in the following for convenience without loss of generality. The channel coefficients $h_{BA},h_{AB},h_{AE},h_{BE}$ are assumed to have a joint probability distribution $P\left(h_{BA},h_{AB},h_{AE},h_{BE}\right)$.
All noise terms $z_A, z_B, z_{E1}, z_{E2}$ are \iid $\rgauss{0}{1}$ distributed. Furthermore, they are independent of all channel coefficients.
From the training signal Bob has sent, Alice obtains $y_A$, which she can use to estimate $h_{BA}$. On the other hand, Bob receives $y_B$, in order to estimate $h_{AB}$. Very often, $h_{BA}$ and $h_{AB}$ are just dependent, but not equal due to non-ideal reciprocity. 
Eve receives the training signal directly from Alice and Bob in $y_{E1}$ and $y_{E2}$, respectively. 

Since reflection at the antenna might severely affect the overall security of the key agreement approach, it is reasonable to extend the basic model. The malicious, but passive eavesdropper Eve intercepts the training signals and, in addition, its reflections. Therefore, Eve obtains two more observations: 
\begin{align}
\label{eq:observations_refl3}
y_{E3} & = \alpha h_{BA}h_{AE}x+z_{E3}  \\
\label{eq:observations_refl4}
y_{E4} & = \alpha h_{AB}h_{BE}x+z_{E4}.
\end{align}
Here, it is assumed that the eavesdropper has the capability of decoupling all four received signals \cref{eq:observations_wo_refl1,eq:observations_wo_refl2,eq:observations_refl3,eq:observations_refl4} without interference, e.g., by using multi-antenna techniques. Some amount of power is reflected at Alice's antenna and re-radiated as new transmitted signal, which can be observed by Eve in $y_{E3}$. The parameter $\alpha$ with $\abs{\alpha}<1$ describes the proportion of Alice's received signal that is reflected. This parameter comprises all contributions to the reflection like the re-radiation, the impedance mismatch and the structural scattering, which are described in more detail in section \ref{sec:sources_reflection}.
 Accordingly Eve's observation $y_{E4}$ stems from the reflection at Bob's antenna. Again the noise terms $z_{E3}, z_{E4}$ are \iid $\rgauss{0}{1}$ distributed.

\section{Sources of reflection}\label{sec:sources_reflection}
\begin{figure}
\centering
\includegraphics{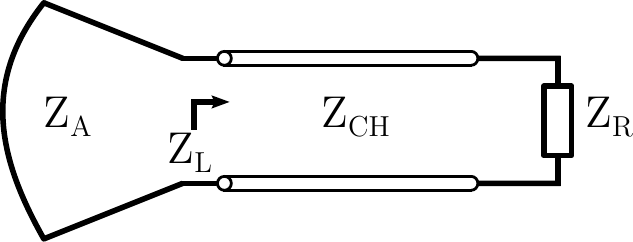}
\caption{Receiver system with antenna impedance $Z_A$, transmission line impedance $Z_{\text{CH}}$, receiver impedance $Z_R$ and overall load impedance $Z_L$ at the feed point \cite{src:best2009}.}
\label{fig:recv}
\end{figure}

From past work on antenna theory \cite{src:balanis2005,src:best2009,src:collin2003}, three possible sources of reflections can be identified.
\begin{enumerate}
\item Some power reflection might result from impedance mismatches in the receiver network. Relevant mismatches can occur in any elements close to the antenna (see figure \ref{fig:recv}). Usually the antenna impedance $Z_A$ is conjugately matched to the load impedance $Z_L$ in order to achieve maximum power throughput. However, optimal power transfer for $Z_A=Z^*_L$ might conflict with the goal of minimizing reflections, which yields $Z_A=Z_L$.
\item The current flow inside the receiving antenna contributes to the reflection. Even with perfectly matched circuits, some amount of power is not fed into the circuit but re-radiated in a scattering field instead.
\item The antenna structure itself introduces a scattering field. This field depends on the structure, shape and material of the antenna and has a different directivity pattern than the re-radiated scattering field.
\end{enumerate}
Scattered and re-radiated portions of power from the receiving antenna are related to a scattering field. In literature, the total scattering field is often described as the superposition of two components
\begin{equation}
\label{equ:scatter_field}
\vec{E}_S=\vec{E}_S(Z_L=\infty)-\frac{V_{\text{OC}}}{Z_L+Z_A}\vec{E}_r.
\end{equation}
The component $\vec{E}_r$ is the scattering field re-radiated by the antenna for unit input current. On the other hand, $\vec{E}_S(Z_L=\infty)$ denotes the scattering field in case of an open-loop circuit at the antenna feed point. Furthermore, $V_{\text{OC}}$ the induced voltage at the antenna feed point, $Z_A$ the antenna impedance and $Z_L$ the complex load impedance. The antenna impedance is obtained by
\begin{equation}
Z_A = R_l + R_r +\text{j} X_A,
\end{equation} 
where $R_l$ is the loss resistance, $R_r$ the radiation resistance and $X_A$ the antenna reactance.
\begin{figure}
\centering
\includegraphics{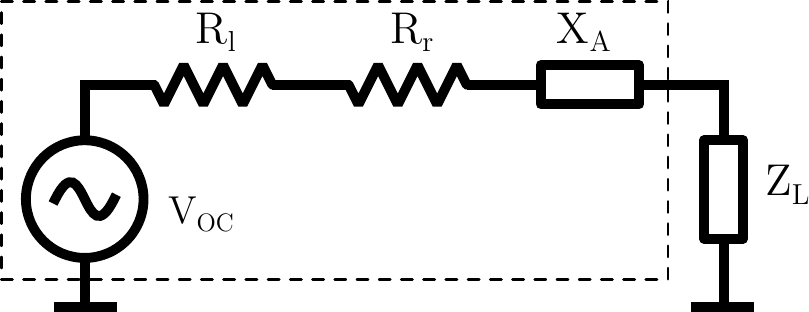}
\caption{Th\'{e}venin equivalent circuit for a receiving antenna.}
\label{fig:thevenin_model}
\end{figure}
A popular model to describe the behaviour of an antenna in reception mode is the Th\'{e}venin model, which is depicted in \figref{fig:thevenin_model}. It is a good approximation at least for dipole antennas which operate close to the resonance frequency. The model assumes the same antennas at transmitter and receiver (full reciprocity) which are located in far field with respect to each other. In addition the antennas should be small, the dielectric without loss and the ground ideal. As depicted in \figref{fig:thevenin_model}, the model basically characterizes a series RLC circuit. A current flowing through the radiation resistance $R_r$ contributes to the scattering field of the antenna, which is basically captured by $\vec{E}_r$ in \eqref{equ:scatter_field}. In order to evaluate the impact of re-radiation in more detail, some power calculations are made on the basis of Th\'{e}venin equivalent circuit. To make following calculations simpler, a perfectly conjugate matched load is assumed.
\begin{equation}
Z_L =Z^*_A = R_l + R_r - \j X_A
\end{equation}
Now the complex power that is fed into the load at the receiver side can be calculated.
\begin{align}
S_L = \frac{1}{2} U_L I^*_L 
& = \frac{1}{8}\frac{Z_L}{\left( R_l + R_r \right)^2}V^2_{OC}
\end{align}
So the power output into the load is
\begin{equation}
\label{eq:power_load}
P_L = \Re{S_L} =\frac{1}{8}\frac{V^2_{OC}}{\left( R_l + R_r \right)},
\end{equation}
while the power dissipated as heat is
\begin{equation}
\label{eq:power_diss}
P_l = \frac{1}{8}\frac{R_l}{\left( R_l + R_r \right)^2}V^2_{OC}
\end{equation}
and finally the re-radiated power is
\begin{equation}
\label{eq:power_rera}
P_r = \frac{1}{8}\frac{R_r}{\left( R_l + R_r \right)^2}V^2_{OC}
\end{equation}
Using \cref{eq:power_load,eq:power_diss,eq:power_rera}, the total power at the receiver side is
\begin{equation}
\label{eq:power_total}
P_{total} = P_L + P_l + P_r =\frac{1}{4}\frac{V^2_{OC}}{\left( R_l + R_r \right)}.
\end{equation}
From these considerations, one can define a \textit{re-radiation ratio}
\begin{equation}
\label{eq:reradiation_ratio}
r=\frac{P_r}{P_{total}}=\frac{1}{2}\frac{R_r}{\left( R_l + R_r \right)}
\end{equation}
This straightforward result means that for a completely lossless antenna always half the total power is re-radiated. However, this result does not fully capture the total scattering power since the component $\vec{E}_S(Z_L=\infty)$ from \eqref{equ:scatter_field} is not considered, which is the residual electric field if the feed circuit is open and no current flows. 
While capturing the impact of \eqref{eq:reradiation_ratio} in secret-key generation is a future research topic of itself, here we employ the basic model given in \eqref{eq:observations_refl3} and \eqref{eq:observations_refl4} with a coefficient $\alpha$ to evaluate the essential impact the reflections might have on secret-key rates.
This is further justified by the investigations of \cite{src:dottling2011}, in which it was shown by practical measurements that a significant amount of power is scattered from the receiver side for an antenna in open-loop mode. 
\section{Secret-key rates}
\label{sec:secret_key_rates}
The overall setup refers to the source-type model with wiretapper of \cite{src:ahlswede1993}, where the legitimate users agree on a secret key. The observations at Alice, Bob and Eve are interpreted as the outputs of memoryless, but correlated sources. The upper bound on the secret key rate is therefore~\cite{src:ahlswede1993}
\begin{align}
\label{eq:sk_up_reflect_init}
\overline{R}_{SK} &= \mui{y_A; y_B|y_{E1},y_{E2},y_{E3},y_{E4}} \nonumber \\
&= \den{y_A|y_{E1},y_{E2},y_{E3},y_{E4}} \nonumber \\
&-\den{y_A|y_B,y_{E1},y_{E2},y_{E3},y_{E4}},
\end{align}
while the lower bound is~\cite{src:maurer1993}
\begin{align}
\label{eq:sk_lo_reflect_init}
\uline{R}_{SK} = \mui{y_A; y_B} \nonumber \\
- \min &\left( \mui{y_A; y_{E1},y_{E2},y_{E3},y_{E4}}, \right.  \nonumber\\
& \left. \mui{y_B; y_{E1},y_{E2},y_{E3},y_{E4}} \right).
\end{align}
In the following, some assumptions of the model from section \ref{sec:system_model} are simplified.
The channel coefficients $h_{BA},h_{AB},h_{AE},h_{BE}$ are assumed to be Gaussian distributed with zero mean and variances $\sigma_{BA}^2, \sigma_{AB}^2,\sigma_{AE}^2,\sigma_{BE}^2$, respectively.
Furthermore, $h_{BA}$, $h_{AB}$ are independent of $h_{AE}$ and $h_{BE}$. This assumption is justified if Eve is located further away from Alice and Bob such that the channel coefficients undergo a rapid spatial decorrelation. As a consequence, the observations $y_{E1}$ from \eqref{eq:observations_wo_refl1} and $y_{E2}$ from \eqref{eq:observations_wo_refl2} can be neglected for all rate calculations since they are independent of all other observations. The channel coefficients $h_{AB}$ and $h_{BA}$ are correlated with $\rho_{AB}$, while $h_{AE}$ and $h_{BE}$ are correlated with $\rho_{E}$. Furthermore, the variances of the legitimate users' channels are set to $\sigma^2=\sigma_{AB}^2=\sigma_{BA}^2$ and for the eavesdropper's channels to $\sigma_{E}^2=\sigma_{AE}^2=\sigma_{BE}^2$. Conclusively, there are only four observations to consider: 
\begin{align}
\label{eq:siso_real_observations_simplified}
y_A & = h_{BA}+z_A \nonumber \\
y_B & = h_{AB}+z_B \nonumber \\
y_{E3} & = \alpha h_{BA}h_{AE}+z_{E3} \nonumber \\
y_{E4} & = \alpha h_{AB}h_{BE}+z_{E4} 
\end{align} 
The SNR at Alice' and Bob's receiver is
\[ 
\snr=\snra=\snrb=\sigma^2,
\]
while Eve's SNR is equal to
\[ 
\snre=\alpha^2\sigma_E^2\sigma^2 = \alpha^2\sigma_E^2\snr,
\]
since only the reflected observations are considered.

Finally, for comparison, we derive secret-key rates if the reflection is not taken into account. In this case the eavesdropper does not have any observation at all and the secret-key capacity is
\begin{align}
\label{eq:sk_siso_wo_eve}
C_{SK} & = I(y_A; y_B) = h(y_A)+h(y_B)-h(y_A, y_B) \nonumber \\
& = \frac{1}{2}\log_2\frac{\left( 1 + \sigma^2 \right)^2}{\left( 1 + \sigma^2 \right)^2 - \rho_{AB}^2\sigma^4} \nonumber \\
& = \frac{1}{2}\log_2\frac{\left( 1 + \snr\right)^2}{\left( 1 + \snr \right)^2 - \rho_{AB}^2 \snr^2} 
\end{align}

The next sections concern the question of deriving closed-form solutions for the bounds of \eqref{eq:sk_up_reflect_init} and \eqref{eq:sk_lo_reflect_init}.
\section{Key rates without CSIE}
\label{sec:rates_without_csie}
In this section, Eve has no CSIE (channel state information at the eavesdropper), i.e. the channel coefficients $h_{AE}$ and $h_{BE}$ are unknown to her.  
The upper bound on the secret-key rate is   
\begin{align}
\label{eq:sk_up_bound_1}
\overline{R}_{SK} &= \mui{y_A; y_B|y_{E3},y_{E4}} \nonumber\\
&= \den{y_A|y_{E3},y_{E4}}-\den{y_A|y_B,y_{E3},y_{E4}} \\
\label{eq:sk_up_bound_2}
&= h(y_A,y_{E3},y_{E4})+h(y_B,y_{E3},y_{E4}) \nonumber\\
&\quad -h(y_A, y_B,y_{E3},y_{E4})-h(y_{E3},y_{E4})
\end{align}
while the definition of mutual information is used in \eqref{eq:sk_up_bound_1} and the chain rule of entropies in \eqref{eq:sk_up_bound_2}. The lower bound is accordingly 
\begin{align}
\label{eq:sk_lo_bound_1}
\uline{R}_{SK} &= \mui{y_A; y_B} -  \nonumber\\
&\quad\min\left( \mui{y_A; y_{E3},y_{E4}}, \mui{y_B; y_{E3},y_{E4}} \right) \nonumber \\
&= \den{y_A} + \den{y_B} - \den{y_A, y_B} - \den{y_{E3}, y_{E4}} \nonumber \\
& \quad - \min\{( \den{y_A}-\den{y_A,y_{E3},y_{E4}}, \nonumber \\
& \quad \den{y_B}-\den{y_B,y_{E3},y_{E4}} )\}  \\
\label{eq:sk_lo_bound_2}
&= \den{y_B} - \den{y_A, y_B} - \nonumber \\
& \quad \den{y_{E3}, y_{E4}} + \den{y_A,y_{E3},y_{E4}}
\end{align}
while in \eqref{eq:sk_lo_bound_1} the chain rule of entropies is applied again and \eqref{eq:sk_lo_bound_2} exploits the fact that $y_A$ and $y_B$ are identically distributed. Therefore the entropies of $y_A$ and $y_B$ are the same and \eqref{eq:sk_lo_bound_1} takes the minimum of equal values.
   
Since the multivariate distributions from \eqref{eq:sk_lo_bound_1} and \eqref{eq:sk_lo_bound_2} are non-Gaussian due to the reflection terms $y_{E3}$ and $y_{E4}$, we use entropy estimators to calculate the joint entropies.

\begin{figure}
\centering
	\centering 
	\includegraphics[width=0.47\textwidth]{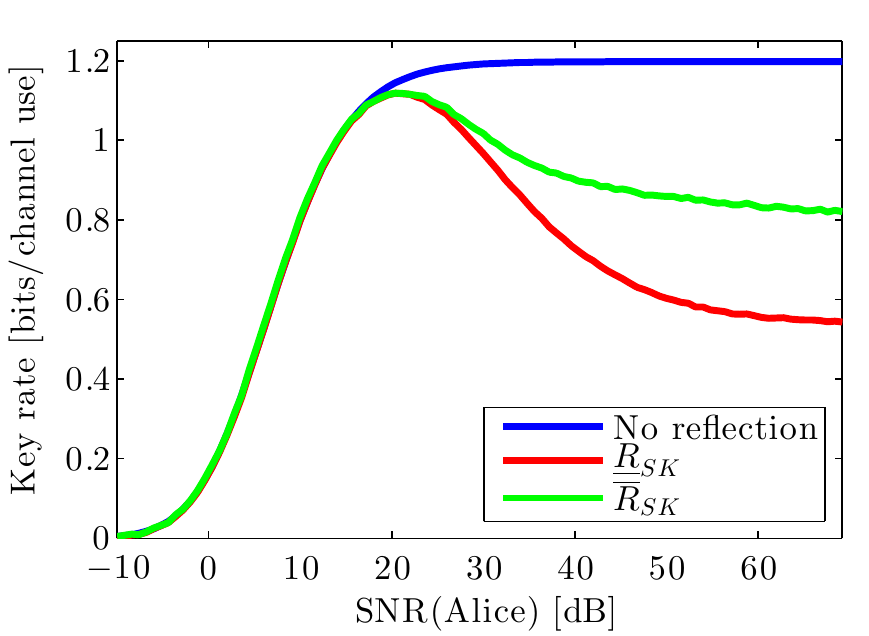}
	\caption{Estimated bounds of reflection model without CSIE and parameters $\rho_{AB}=0.9$, $\alpha=0.05$, $\rho_{E}=0.1$.}
	\label{fig:keyrate_snr_reflect_wocsie}
\end{figure}

The entropy estimation utilizes a combination of the $k$-nearest neighbor estimator (NNE) from \cite{src:kraskov2004} and a kernel density estimator (KDE) from~\cite{src:gray2003}.
The results of the simulation for the special case $\sigma_E^2=1$ are illustrated in \figref{fig:keyrate_snr_reflect_wocsie}. This case covers the situation when the power of the training symbols is increased by the legitimate users in order to have improved SNR, but Eve benefits from that as well.
The simulation reveals that, for relatively small SNR, the key rate is almost unaffected by Eve's presence. However, at a certain SNR, the lower and upper bound diverge and decrease. Furthermore it turns out that the parameter $\alpha$ determines the width of the lower bound's maximum. The parameter basically has a ``retarding'' effect on the point when the rate drops, so for a small $\alpha$ the maximum broadens. In addition, the parameter $\rho_{AB}$ defines the saturation level for high SNR. 
Surprisingly even with negligible noise at the eavesdropper, a positive secret-key rate is still possible. This might be due to the fact that Eve always encounters her unknown channel coefficients, which impede her in the estimation of $h_{AB}$ or $h_{BA}$.

\begin{figure}
	\centering
	\includegraphics[width=0.47\textwidth]{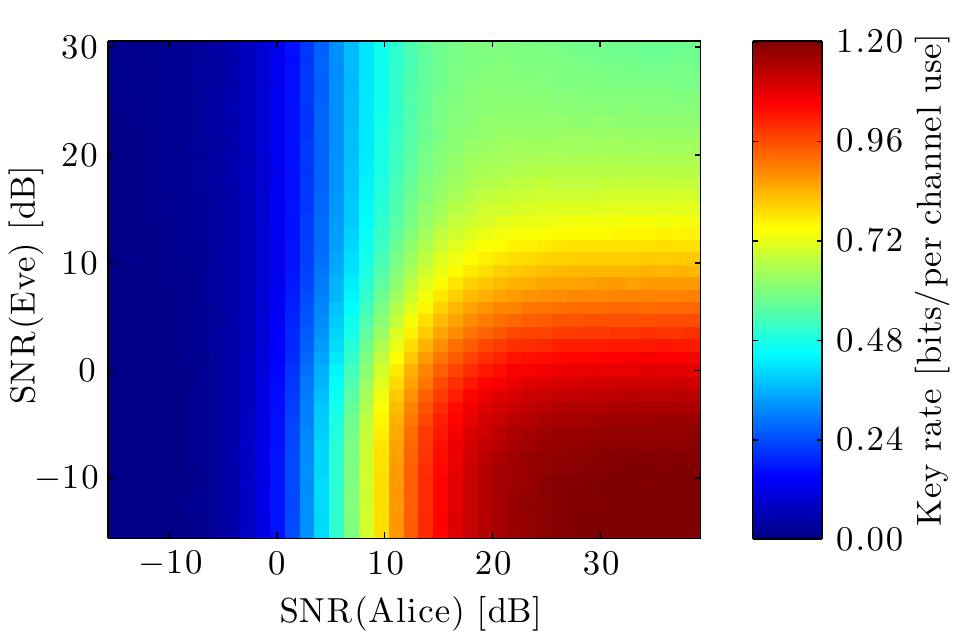}
	\caption{Achievable key rates of the reflection model without CSIE and parameters $\rho_{AB}=0.9$, $\alpha=0.05$, $\rho_{E}=0.1$. The values at the colorbar denote the key rates in bits per channel use.}
	\label{fig:keyrate_snr_reflect_3d_wocsie}
\end{figure}

If we consider the more general case, i.e. if $\sigma^2_E$ can be chosen arbitrarily, the lower bound is depicted in \figref{fig:keyrate_snr_reflect_3d_wocsie}.
For low SNR at Alice's receiver, no secret-key agreement is possible. However, for higher SNR, there is always a positive secret-key rate regardless of how accurate Eve's observation might be. 
The lower bound of \figref{fig:keyrate_snr_reflect_wocsie} is retrieved as a special case, if one cuts through the plane of \figref{fig:keyrate_snr_reflect_3d_wocsie} in a line parallel to the main diagonal with $\snre(\text{dB}) = \snra(\text{dB})+20\log_{10}(\alpha)$.
\section{Key rates with CSIE}
\label{sec:rates_with_csie}
\begin{figure}
	\centering 
	\includegraphics[width=0.47\textwidth]{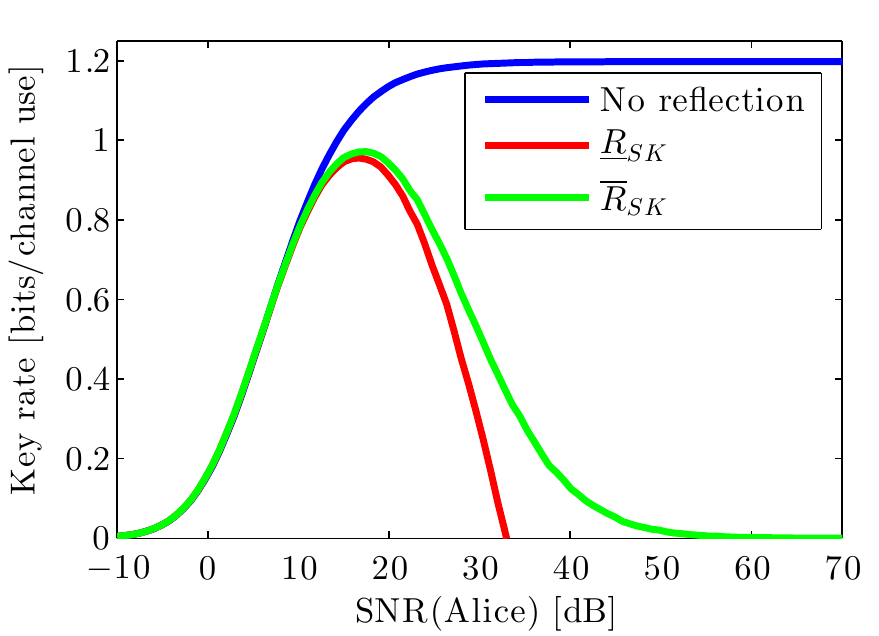}
	\caption{Secret-key rates with full CSIE and parameters $\rho_{AB}=0.9$, $\alpha=0.05$, $\rho_{E}=0.1$.}
	\label{fig:keyrate_snr_reflect_csie_1}
\end{figure}

\begin{figure}
	\centering 
	\includegraphics[width=0.47\textwidth]{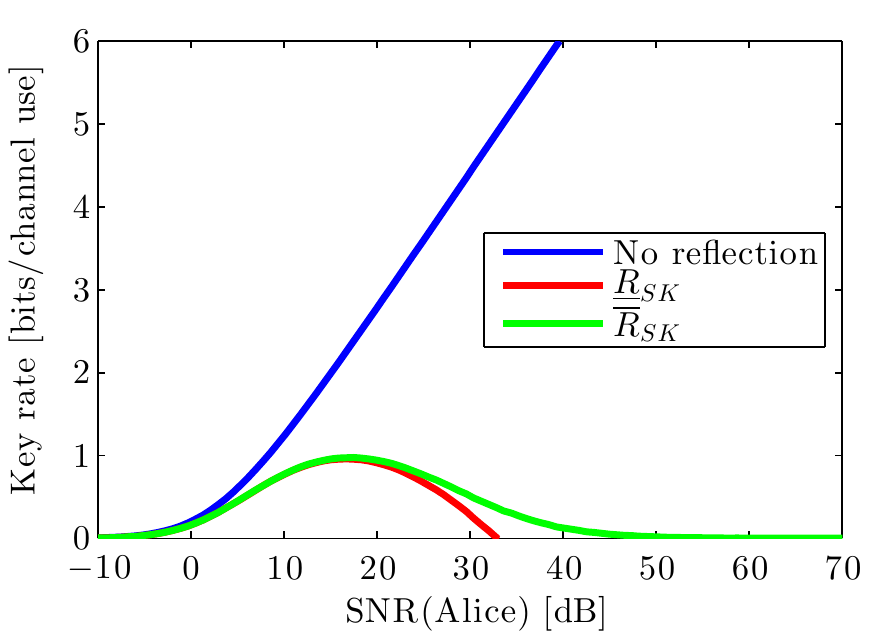}
	\caption{Secret-key rates with full CSIE and parameters $\rho_{AB}=1$, $\alpha=0.05$, $\rho_{E}=0.1$.}
	\label{fig:keyrate_snr_reflect_csie_2}
\end{figure}

\begin{figure}
	\centering
	\includegraphics[width=0.47\textwidth]{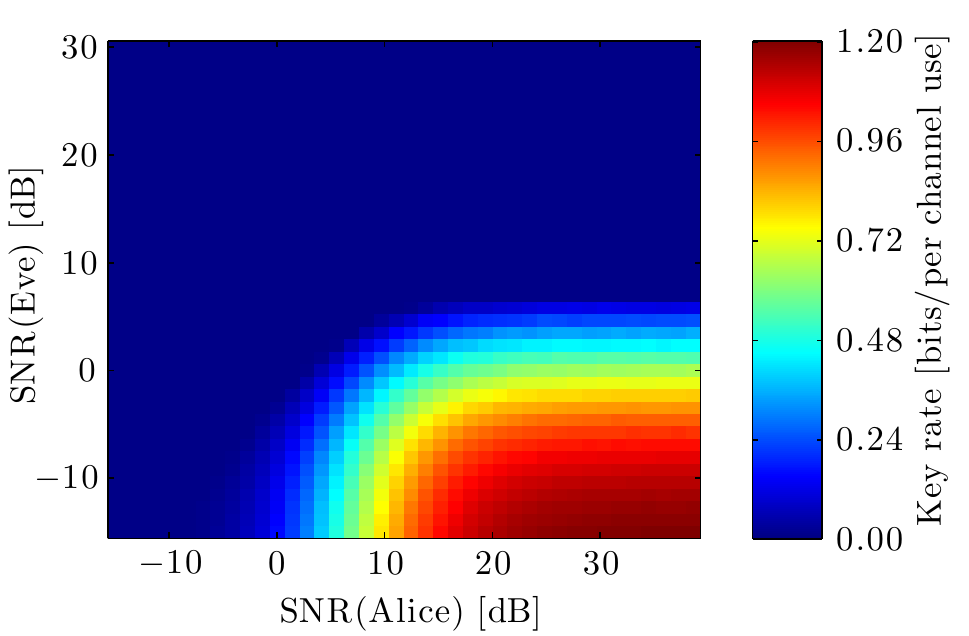}
	\caption{Achievable key rates of the reflection model with CSIE and parameters $\rho_{AB}=0.9$, $\alpha=0.05$, $\rho_{E}=0.1$. The values at the colorbar denote the key rates in bits per channel use.}
	\label{fig:keyrate_snr_reflect_3d_csie}
\end{figure}

Another scenario comes into play if Eve knows her channel coefficients perfectly, thus she has full CSIE. The bounds of \eqref{eq:sk_up_reflect_init} and \eqref{eq:sk_lo_reflect_init} are then conditioned on $h_{AE}$ and $h_{BE}$, respectively. As a consequence, the model from \eqref{eq:siso_real_observations_simplified} simply reduces to the multivariate Gaussian case.
Therefore, one need to average over all possible realizations of the channel coefficients in order to calculate average rates. 
The upper bound is obtained by
\begin{align}
\overline{R}_{SK} &= \mui{y_A; y_B|y_{E3},y_{E4},h_{AE},h_{BE} }\nonumber \\
&= E_{h_{AE},h_{BE}} 
\left\{ I\left\{y_A;y_B|y_{E3},y_{E4}, \right.\right.\nonumber \\
&\quad h_{AE}=\hat{h}_{AE},h_{BE}=\hat{h}_{BE} \}\}.
\end{align}
and the lower bound by
\begin{alignat}{2}
\underline{R}_{SK} &= \mui{y_A; y_B}& &- \min\left\lbrace\mui{y_A;y_{E3},y_{E4},h_{AE},h_{BE} }, \right. \nonumber\\
\label{eq:lb_wocsie_1}
 & & &\left.\mui{y_B; y_{E3},y_{E4},h_{AE},h_{BE} }\right\rbrace\\
&= \mui{y_A; y_B}& &- \mui{y_A;h_{AE},h_{BE}} \nonumber \\
\label{eq:lb_wocsie_2}
& & &-  \mui{y_A;y_{E3},y_{E4}|h_{AE},h_{BE} } \\
&= \mui{y_A; y_B}& &- E_{h_{AE},h_{BE}} 
\left\{ I\left\{y_A;y_{E3},y_{E4}| \right.\right.\nonumber \\
\label{eq:lb_wocsie_3}
& & & h_{AE}=\hat{h}_{AE},h_{BE}=\hat{h}_{BE} \}\},
\ 
\end{alignat}
where in \eqref{eq:lb_wocsie_1} the minimum of equal terms of mutual information is taken since they both depend on random variables with the same distribution. In \eqref{eq:lb_wocsie_2} the chain rule of mutual information is utilized and \eqref{eq:lb_wocsie_3} exploits the fact that $h_{AE}$ and $h_{BE}$ are independent of Alice's observation. The result is illustrated in \figref{fig:keyrate_snr_reflect_csie_1} and in addition in \figref{fig:keyrate_snr_reflect_csie_2} for the case of full reciprocity.   
Compared to the case without CSIE, both bounds are dropping to zero when Eve's SNR increases. Thus full CSIE means the worst-case scenario in terms of secret-key rate. More specific, the terminals obtain the observations of \eqref{eq:siso_real_observations_simplified}, but now the channel coefficients $h_{AE}$ and $h_{BE}$ are fully known by the eavesdropper. The resulting achievable key rates are illustrated in \figref{fig:keyrate_snr_reflect_3d_csie}. 
Now the saturation level has vanished for higher values of $\snre$. Naturally Eve obtains a profound knowledge of the channel coefficient for high SNR, therefore no positive secret-key rate is possible.

\section{conclusion}
\label{sec:conclusion}
In this paper, the impact of reflections, a phenomenon present at any practical antenna, on the secret-key rate based on the randomness of the wireless channel was investigated. At first, the rather optimistic case without channel state information at the eavesdropper (CSIE) was considered. At low SNR, the secret-key rates are almost unaffected by eavesdropping. From a certain point on, the key rates drop due to Eve's observations, depending on the antenna reflection. At high SNR the key rate might still be positive since Eve is impeded by the unknown channel realizations. The saturation level at high SNR is mostly determined by the correlation of Alice's and Bob's observations. Considering the more realistic case of full CSIE, however, the lower bound drops significantly without a saturation level at high SNR since Eve learns the information on the key error-free due to the reflections from the antennas of the legitimate nodes. As a consequence, there exists an optimal SNR for which the secret key rate is maximal and otherwise is less due to the effect of additive noise in the communication chain or the reflections at the antennas.


\bibliographystyle{IEEEtran}
\bibliography{IEEEabrv,references}

\end{document}